%% file: subtyping-coraglia-emmenegger.tex
\documentclass[a4paper,UKenglish,cleveref,autoref]{amsart}

\usepackage{hyperref}
\usepackage{quiver}
\usepackage{todonotes}
\usepackage{bussproofs}
\usepackage{tabularx}
\usepackage{tensor}

\newcommand{\ty}{\, \texttt{Type}}

\newcommand{\ie}{\textit{i.e.}\ }

\newcommand{\catof}[1]{\mathbf{#1}}
\newcommand{\ctg}[1]{\mathcal{#1}}
\newcommand{\cod}{\mathrm{cod}}
\newcommand{\dom}{\mathrm{dom}}
\newcommand{\id}{\mathrm{id}}
\newcommand{\Id}{\mathrm{Id}}
\newcommand{\opp}{^{\mathrm{op}}}
\newcommand{\due}{^{\to}}

\newcommand{\vcomma}[2]{#1 \!\downarrow^v\! #2}
\newcommand{\Cat}{\catof{Cat}}
\newcommand{\Set}{\catof{Set}}

\newcommand{\Ty}{\text{Ty}}
\newcommand{\Tm}{\text{Tm}}
\newcommand{\Fib}{\catof{Fib}}
\newcommand{\Psd}{\catof{Psd}}
\newcommand{\duu}{\dot{\ctg{U}}}
\newcommand{\uu}{\ctg{U}}
\newcommand{\du}{\dot{u}}
\newcommand{\ctx}{\ctg{B}} 

\newcommand{\ee}{\ctg{E}}

\newcommand{\dv}{\dot{v}}

\newcommand{\bb}{\ctg{B}}
\newcommand{\gcwf}{\catof{GCwF}}
\newcommand{\one}{\mathsf{1}}

\newcommand{\funty}{\mathsf{Fun}}
\newcommand{\absty}{\mathsf{abs}}
\newcommand{\vopp}{^{\mathrm{o}}}
\newcommand{\nn}{\mathbb{N}}

\input{./header.tex}
\usepackage{lmodern}
\renewcommand{\autoref}{\Cref}
\begin{document}
\title{Categorical models of subtyping}
\author[Coraglia]{Greta Coraglia}
\email{greta.coraglia@unimi.it}
\author[Emmenegger]{Jacopo Emmenegger}
\email{emmenegger@dima.unige.it}

\maketitle

\bibliographystyle{alpha}

\begin{abstract}
Most categorical models for dependent types have traditionally been heavily \textit{set} based: contexts form a category, and for each we have a set of types in said context -- and for each type a set of terms of said type. This is the case for categories with families, categories with attributes, and natural models; in particular, all of them can be traced back to certain discrete Grothendieck fibrations. We extend this intuition to the case of general, non necessarily discrete, fibrations, so that over a given context one has not only a set but a \textit{category} of types.

We argue that the added structure can be attributed to a notion of subtyping that shares many features with that of \emph{coercive} subtyping, in the sense that it is the product of thinking about subtyping as an abbreviation mechanism: we say that a given type $A'$ is a subtype of $A$ if there is a unique coercion from $A'$ to $A$. Whenever we need a term of type $A$, then, it suffices to have a term of type $A'$, which we can `plug-in' into $A$.

For this version of subtyping we provide rules, coherences, and explicit models, and we compare and contrast it to coercive subtyping as introduced by Z. Luo and others. We conclude by suggesting how the tools we present can be employed in finding appropriate rules relating subtyping and certain type constructors.
\end{abstract}

\tableofcontents

The notion of subtyping is often quite tricky because of its double nature: on one hand, it is meant to represent a relation so simple that one would like to not be bothered to look too much into it but, at the same time, programming languages do not do well with things that are left unsaid. In \cite[$\S$15.1]{citeulike:105547}, subtyping $A\leq B$ is defined as a relation between two types $A$ and $B$ such that, if a term of type $A$ is provided, than it can be `safely' used in a context where $B$ is expected, and the program should not falter if this happens. This is often called the \emph{principle of safe substitution}, and is encoded via a new typing rule that goes by the name of \emph{subsumption}.
\[
(Sub)\frac{\Gamma\vdash a:A\quad\Gamma \vdash A\leq B}{\Gamma \vdash a:B}
\]
\vspace{.0001em}

\noindent The problem with subsumption is that it threatens to break certain structures and properties such as canonicity or induction principles. Many variants of subtyping have been proposed, and one in particular has proved itself to be well-behaved with respect to such issues, and that is \emph{coercive subtyping} \cite{10.1093/logcom/9.1.105}.

In the present work we tackle the problem of subtyping from the point of view of its categorical semantics: we extend a known model to naturally include a notion of subtyping (\autoref{sec:1}), and show how this turns out to be in fact quite closely related to coercive subtyping (\autoref{sec:2}). We then study some examples and applications (\autoref{sec:3}), and conclude by observing some properties that that this new notion intrinsically shows (\autoref{sec:4}).

\section{Categorical models of dependent types}\label{sec:1}
The relation between type theory and category theory is one which has been widely studied, and has in the years produced a large variety of models and structures. The interest of the present work is in those that strongly employ objects and techniques coming from the theory of Grothendieck fibrations.

\subsection{Grothendieck fibrations}\label{intro_to_fib}
Fibrations were introduced by A. Grothendieck \cite{grothendieck_fibdesc_SGA1} for purposes pertaining algebraic geometry, but were soon found extremely useful to describe certain phenomena in logic \cite{lawvere1963functorial}. We will recall key results and definitions when needed, but we refer to \cite[Chapter 8]{BOR2} for a detailed introduction.

\begin{definition}[Cartesian morphism]
Let $p\colon\ee\to\bb$ a functor and \mbox{$s\colon B\to A$} a morphism in $\ee$. We say that $s$ is \emph{$p$-cartesian} or \emph{cartesian} over $\sigma\colon \Theta\to\Gamma$ if $p(s)=\sigma$ and for any other $r\colon C\to A$ and $\tau$ such that $p(r)=\sigma\circ \tau$ there is a unique $t\colon C\to B$ in $\ee$ such that $p(t)=\tau$ and $s\circ t =r $.
\begin{equation}\label{cartesian}
\begin{tikzcd}[ampersand replacement=\&]
	C \\
	\& B \& A \&\& \ee \\
	\Xi \\
	\& \Theta \& \Gamma \&\& \bb
	\arrow["s"', from=2-2, to=2-3]
	\arrow["r", curve={height=-6pt}, from=1-1, to=2-3]
	\arrow["\sigma"', from=4-2, to=4-3]
	\arrow["\sigma\circ\tau", curve={height=-6pt}, from=3-1, to=4-3]
	\arrow["\tau"', from=3-1, to=4-2]
	\arrow["p", from=2-5, to=4-5]
	\arrow[dashed, from=1-1, to=2-2]
\end{tikzcd}
\end{equation}
We say that $s$ is a \emph{($p$-)cartesian lifting} of $\sigma$, and it is often denoted $s_{A,\sigma}\colon\sigma^*A\to A$.
\end{definition}
\begin{definition}[Fibration]\label{defn:fibration}
A functor $p\colon\ee\to\bb$ is a \emph{fibration} if for all $A$ in $\ee$, each $\sigma:\Theta \to pA$ has a cartesian lifting. We also say that $\ee$ is \emph{fibered over} $\bb$ or that $\ee$ is \emph{over} $\bb$. Oftentimes $\bb$ is called the \emph{base category} and $\ee$ the \emph{total category} of $p$.
\end{definition}
From the point of view of the logic, we can consider $\bb$ to be a category of contexts and substitutions/terms, and $\ee$ to be a category of formulae/types, so that each formula is sent to its corresponding context. Asking that for each term and formula there exist a lifting, then, amounts to asking that substitution can be computed and that it indeed produces a formula in then new context. A classical example is that of the Lindenbaum-Tarski algebra for a given first-order theory (\autoref{lt-fibration}).

A particular feature of fibrations is that they induce a factorization system on their total category, because for any given $r\colon C\to A$, we can always instantiate the diagram in (\ref{cartesian}) as follows,
\[\begin{tikzcd}[ampersand replacement=\&]
	C \\
	{\sigma^*A} \& A \&\& \ee \\
	\Theta \\
	\Theta \& \Gamma \&\& \bb
	\arrow["{s_{A,\sigma}}"', from=2-1, to=2-2]
	\arrow["r", curve={height=-6pt}, from=1-1, to=2-2]
	\arrow["\sigma"', from=4-1, to=4-2]
	\arrow["\sigma", curve={height=-6pt}, from=3-1, to=4-2]
	\arrow["\id"', from=3-1, to=4-1]
	\arrow["p", from=2-4, to=4-4]
	\arrow[from=1-1, to=2-1]
\end{tikzcd}\]
producing a factorization of $r$. Maps in the left class of such factorization system, meaning maps that are sent to identities, are called \emph{vertical}.
\begin{proposition}[Vertical/cartesian factorization system]\label{vertcart_facsys}
Consider \hbox{$p\colon\ee\to\bb$} a fibration. The classes of vertical and cartesian morphisms form a orthogonal factorization system on $\ee$. It additionally has the following properties:
\begin{enumerate}
\item if $g$ and $gf$ are vertical, then so is $f$;
\item pullbacks of vertical maps along cartesian ones exist and are vertical.
\end{enumerate}
\end{proposition}
If cartesian maps can be thought of as substitutions, vertical maps relate formulae/types in the same context precisely in a way that `does nothing' to the underlying terms.

Not only are vertical maps part of a factorization system, but they can be shown to form a category: more precisely, for each context $\Gamma$ in $\bb$, one can define a category $\ee_\Gamma$ having for objects those in $\ee$ that are sent to $\Gamma$, and for morphisms those in $\ee$ that are vertical over $\id_\Gamma$. This intuition is part of one of the most meaningful results of the theory, again due to Grothendieck.
\begin{theorem}[\cite{grothendieck_fibdesc_SGA1}]\label{groth_equivalence}
There exists a 2-equivalence
\[
\Fib(\bb)\cong \Psd[\bb\opp,\Cat]
\]
between the 2-category of fibrations (with base $\bb$), functors preserving cartesian maps, and natural transformations, and that of contravariant pseudofunctors (from $\bb$), pseudonatural transformations, and modifications.
\end{theorem}
From left to right, a fibration is sent to a (pseudo)functor that computes for each $\Gamma$ its \emph{fiber} $\ee_\Gamma$. From right to left, it performs what is called the \emph{Grothendieck construction} of a (pseudo)functor, which to an $F\colon\bb\opp\rightsquigarrow\Cat$, maps the fibration $p\colon\int F\to\bb$, where objects of $\int F$ are pairs $(\Gamma,A)$ with $\Gamma$ in $\bb$ and $A$ in $F(\Gamma)$.

Relevant classes of fibrations are those that correspond to functors, which are called \emph{split}, those whose fibers are preorders, which are called \emph{faithful}, and those whose fibers are sets, which are called \emph{discrete}.

\subsection{Categories with families}\label{subsec:cwfs}
\begin{definition}[Category with families, {\cite{dybjer_inttt}}]\label{cwf}
A \emph{category with families (cwf)} is the data of
\begin{itemize}
\item a small category $\bb$ with terminal object $\top$;
\item a functor $\Ty\colon\bb\opp\to\Set$;
\item a functor $\Tm\colon\int\Ty\opp\to\Set$
\item for each $\Gamma$ in $\bb$ and $A$ in $\Ty(\Gamma)$ an object $\Gamma.A$ in $\bb$, together with two projections $\mathsf{p}_A\colon\Gamma.A\to\Gamma$ and $\mathsf{v}_A\in\Tm(\Gamma.A,\Ty\,\mathsf{p}_A (A) )$ such that for each $\sigma\colon\Theta\to\Gamma$ and $a\in\Tm(\Ty\sigma (A))$ there exists a unique morphism $\Theta\to\Gamma.A$ making the obvious triangles commute.
\end{itemize}
\end{definition}
In particular,
$\Ty(\Gamma)$ is a \emph{set}, the set of types in context $\Gamma$, and $\Tm(\Gamma,A)$ is again a set, the set of terms of type $A$ in context $\Gamma$. The terminal object is meant to model the empty context, while the last condition is what provides context extension -- since we are about to give a simpler, equivalent form of it, we do not dwell on it any longer.


Following \autoref{groth_equivalence}, we can turn \autoref{cwf} upside down. What we get is the (equivalent, see \cite[Prop. 1.2]{awodey_2018}) notion of natural model.

\begin{definition}[Natural model, {\cite{awodey_2018}}]\label{natmod}
A \emph{natural model} is the data of
\begin{itemize}
\item a small category $\bb$ (with terminal object $\top$);
\item a discrete fibration $u\colon\uu\to\bb$;
\item a discrete fibration $\du\colon\duu\to\bb$;
\item a fibration morphism $\Sigma\colon\du\to u$ with a right adjoint functor.
\end{itemize}
\end{definition}
In this case $u$ collects types and $\du$ collects terms, as both are fibered on contexts.
\begin{center}
\begin{tabular}{cc}
$\Gamma \vdash A\ty$ & $\Gamma \vdash a: A$\\
$u(A)=\Gamma$ & $\du(a)=\Gamma,\Sigma (a) =A$\\
$u\;\text{in }\Fib^{disc}(\ctx)$ & $\du\;\text{in }\Fib^{disc}(\ctx)$
\end{tabular}
\end{center}
The fibration morphism (\ie a functor making the diagram below commute) $\Sigma$ maps to each term its type, while $\Delta$ computes for each $A$ what in \autoref{cwf} is called $\mathsf{v}_A$, meaning a variable in $A$ `transported' to the context $\Gamma.A$.
\begin{gather}\label{rules-gcwf}
\begin{tikzcd}[ampersand replacement=\&]
	\duu \&\& \uu \\
	\& \bb
	\arrow["u", from=1-3, to=2-2]
	\arrow["\du"', from=1-1, to=2-2]
	\arrow[""{name=0, anchor=center, inner sep=0}, "\Sigma"', curve={height=6pt}, from=1-1, to=1-3]
	\arrow[""{name=1, anchor=center, inner sep=0}, "\Delta"', curve={height=6pt}, from=1-3, to=1-1]
	\arrow["\dashv"{anchor=center, rotate=90}, draw=none, from=0, to=1]
\end{tikzcd}
\qquad
(\Sigma)\frac{\Gamma\vdash a:A}{\Gamma\vdash A\ty}
\quad
(\Delta)\frac{\Gamma\vdash A\ty}{\Gamma.A\vdash \mathsf{v}_A:A}
\end{gather}

\subsection{Generalized categories with families}
Our intuition, now, is that we want to use the theory of fibrations to generalize categories with families (or natural models) to the case where fibers over a given context are no longer a set, but a (small)  category. If we manage to do this swiftly, we will have found a model for dependent types that introduces and additional relation between types in the same context, so that this relation does imply absolutely no substitution. This leads us to the following definition.

\begin{definition}[Generalized category with families, {\cite{cjd,coraglia_phd}}]\label{gcwf}
A \emph{generalized category with families (gcwf)} is the data of
\begin{itemize}
\item a small category $\bb$ (with terminal object $\top$);
\item a fibration $u\colon\uu\to\bb$;
\item a fibration $\du\colon\duu\to\bb$;
\item a fibration morphism $\Sigma\colon\du\to u$ with a right adjoint functor, and unit and counit with cartesian components.
\end{itemize}
\end{definition}
Notice that the adjoint pair in \autoref{gcwf} is \emph{not} fibered, as $\Delta$ does not make the desired triangle commute and unit and counit have cartesian components.

Of course a gcwf with discrete $u,\du$ is a regular cwf, so that a gcwf is simply a generalization of a well-known model. We choose the name as to remark that this new structure falls into a long tradition of models, which categories with families is perhaps one of the most prominent exponents of, but these could be very easily called `generalized natural models', or something entirely different (cf. \cite{cjd}). We conclude this section with one last result relating gcwfs to other notable structures, namely comprehension categories \cite{comprehensioncats}, as to show that our path did not stray much away from known territory.

\begin{theorem}[\cite{2dimcomp,coraglia_phd}]\label{biequiv_thm}
There is a biequivalence between (the 2-category of) gcwfs and (the 2-category of) comprehension categories.
\end{theorem}

A generalized category with families, then, is precisely \emph{as good a model as} a comprehension category. The purpose of introducing the new structure, though, lies in the clarity of its use, and we hope that the rest of the present work will be a witness to that.

\begin{remark}\label{what_we_skip}
In our exposition we put aside two elements in the discussion on categorical models for dependent types. The first is of course the issue of coherence: since fibrations involve \emph{pseudo}functors, equations for identities and, especially, associativity only hold up to vertical isomorphism, while it is usually preferable to have substitution `on the nose'. If one so wishes, the reader is invited to only use \emph{split} fibrations, meaning those that correspond to strict functors, and morphisms that preserve the splitting. A thorough discussion on the topic can be found in  \cite[\S3]{streicher2022fibred}.

Another sensitive topic is that of definitional equality. For the moment, having other purposes in mind, we settle for interpreting it as identity of objects (in the category over each context) and, therefore, rarely make explicit coherence rules involving it, as they are all trivial from our definitions.
\end{remark}

\section{Vertical maps induce a notion of subtyping}\label{sec:2}
The idea that vertical maps could nicely relate types in the same context is of course not new. The most closely related precedent to this is perhaps \cite{10.1145/2775051.2676970}, followed by the notes in \cite{zeilberger_typeref}, where, quite radically, functors were interpreted to \emph{be} type refinement systems. Our work is similar in spirit -- though one should really be careful of the difference between ``type refinement'' and ``subtyping'' \cite[\S2.2]{zeilberger_typeref} -- but of course based on fibrations instead on functors, hence with a special focus on substitution.

\subsection{Two new judgements}\label{new_judgements}
Now that we are all set, let us describe what it is that we can say in a gcwf, that was not already available in the discrete case. To the two judgements pertaining types and terms, we add two new ones involving subtyping.
\begin{center}
\begin{tabular}{cccc}
$\Gamma \vdash A\ty$ & $\Gamma \vdash a: A$ & $\Gamma \vdash A'\leq_f A$ & $\Gamma \vdash a:_g A$ \\
$u(A)=\Gamma$ & $\du(a)=\Gamma,\Sigma (a) =A$ & $f\colon A'\to A$, $u(f)=\id_\Gamma$ & $g\colon \Sigma a\to A$, $u(g)=\id_\Gamma$
\end{tabular}
\end{center}
In particular, when a notion of subtyping is introduced, it automatically entails a notion of ``sub-typing'' for terms as well. We read the new judgements, respectively, as \emph{$A'$ is a subtype of $A$, as witnessed by $f$} and \emph{$a$ is a term of type a subtype of $A$, as witnessed by $g$}. Recall that vertical maps enjoy some nice properties (\autoref{vertcart_facsys}), and their combination provides us with structural rules for the new judgements.

\begin{proposition}[Structural subtyping rules]\label{strrules}
The following rules are satisfied by a gcwf.
\begin{equation*}
\AxiomC{$\Gamma\vdash a:_g A'$}\AxiomC{$\Gamma\vdash A'\leq_f A$\LeftLabel{(Sbsm)}}\BinaryInfC{$\Gamma\vdash a:_{fg} A$}
\DisplayProof
\qquad
\AxiomC{$\Gamma\vdash A' \leq_f A$}\AxiomC{$\Gamma\vdash A''\leq_g A'$}\LeftLabel{(Trans)}\BinaryInfC{$\Gamma\vdash A'' \leq_{fg} A$}
\DisplayProof
\end{equation*}
\begin{equation*}
\AxiomC{$\Gamma.A\vdash B' \leq_f B$}\AxiomC{$\Gamma\vdash a:_g A$}\LeftLabel{(Sbst)}\BinaryInfC{$\Gamma\vdash B'[a] \leq_{\du(\Delta g \eta_a)^* f} B[a]$}
\DisplayProof
\qquad
\AxiomC{$\Gamma\vdash A' \leq_f A$}\AxiomC{$\Gamma\vdash B\ty$}\LeftLabel{(Wkn)}\BinaryInfC{$\Gamma.B\vdash A' \leq_{(u\epsilon_B)^* f} A$}
\DisplayProof
\end{equation*}
\end{proposition}
Intuitively, Subsumption and Transitivity are both due to the fact that vertical maps compose to vertical maps, while Substitution and Weakening make use of the substitution structure due to the fibration part of the system. We refer to the appendix for further details.

\subsection{Comparison with coercive subtyping}
As we said in the introduction, it turns out that the calculus resulting in this generalization comes close to coercive subtyping. We hope to convince the reader that, despite all the technical differences, they actually entertain the same spirit.

\begin{displayquote}
The basic idea of coercive subtyping is that subtyping is modelled as an abbreviation mechanism: $A$ is a subtype of $B$, if there is a unique coercion $c$ from $A$ to $B$, written as $A<_c B$. Then, if a hole in a context requires an object of type $B$, it is legal to supply an object $a$ of type $A$ – it is equivalent to supplying the object $c(a)$. \cite{LUO201318}
\end{displayquote}
A coercion $c$ from $A$ to $B$ is technically a term of the function type $A\to B$, hence computing $c(a)$ amounts to function application. Coercions are added `manually' to a given type theory, as they are \emph{a priori} not part of the calculus, and the resulting system is shown both to be a conservative extension of the original one and to act well with respect to canonicity.

Let us now collect in a table the main similarities and differences between coercive subtyping and subtyping via vertical maps. We then will consider each point in detail.

\begin{tabularx}{\textwidth}{XX}
&\\
\textit{coercive subtyping} & \textit{categorical subtyping}\\
\hline 
$\Gamma \vdash f: A'\to A$ & $f\colon A'\to A$ vertical over $\Gamma$\\
judgements added to the calculus & judgements `added' the the classical model\\ 
no witnesses for typing judgements & witnesses for typing judgements\\ 
$f$ is unique & $f$ is not necessarily unique\\ 
(Sbsm) via substitution & (Sbsm) via composition\\ 
satisfies (Trans) & satisfies (Trans)\\
satisfies (Sbst) & satisfies (Sbst)\\ 
satisfies (Wkn) & satisfies (Wkn)\\ 
satisfies congruence & satisfies congruence\\
&
\end{tabularx}
The main technical difference is of course what coercions \emph{are}, as on one side they are function terms, on the other they are (particular) morphisms in the total category. In some sense this difference is unavoidable: if one starts from the traditional syntax of a type theory, syntactic objects are all that is available, while if one looks at a very general model as if it \emph{was} a syntax (and one would have some merit in doing so, see \autoref{lt-fibration}), then they have more objects at hand. In both cases, though, subtyping is a notion that is somehow independent of the calculus, because in one case it is added by selecting a choice of witnesses (and adding one new judgement for each), while on the other it is literally \emph{orthogonal} to the rest of the structure.

While we will deal with the matter of uniqueness of coercions in \autoref{faithfulness}, a difference that is unbridgeable is that of dealing with typing judgements: in coercive subtyping, a term can indeed have more than one type, so that a typing judgments is in some sense ambiguous, while in subtyping with vertical maps one can always unambiguously know \emph{by which means} a term is of a certain type, but for this it pays the price of having a whole new set of annotated judgements.

Finally, validity with respect to rules appearing in the table above is common to both systems -- by `congruence', in particular, we mean congruence of the subtyping relation with respect to definitional equality, see \autoref{what_we_skip}. It should be remarked that rules appearing on the `coercive subtyping' side are not all rules required, for example, in \cite{LUO201318}, but they are in a sense the structural ones, as the others aim to discipline either how coercions (as function types) interact with the system, while in our case that is granted.

\subsection{On uniqueness of coercions}\label{faithfulness}

As we suggested in presenting this new perspective on subtyping, uniqueness of coercions is not really an issue. In particular, we have a whole theory of fibrations whose fibers are preorders, we actually call them \emph{faithful} fibrations (cf. \autoref{intro_to_fib}).

Not only that, but we can show that, given a gcwf, if its type fibration is faithful, then so is its term fibration.

\begin{proposition}\label{leftfaith}
Let $(u,\du,\Sigma \dashv \Delta)$ a gwcf. Then $\Sigma$ is a faithful functor.
\end{proposition}

In this sense, faithfulness, hence uniqueness of coercions, can indeed be modeled by simply looking at faithful fibrations, and such a burden is in fact not laid on the choice of the fibration collecting terms.

\begin{remark}[A case against uniqueness]
Though a key feature of coercive subtyping, avoiding uniqueness might have its merits. Consider for example the case of sum types: it seems like one should have two different witnesses for the judgement $\Gamma\vdash A\leq A+A$, one per coproduct injection.
\end{remark}

\subsection{Comparison with related variants of type theory}\mbox{}\\

\noindent\textit{Directed type theory.}
Extending the paradigm connecting types, $\omega$-groupoids, and homotopy theory, to the directed case -- meaning involving $\omega$-categories and directed homotopy theory, one encounters the notion of directed type theory. The underlying intuition wishes to add, for each pair of terms of the same type (and substitutions), a new \emph{asymmetric} `identity' type of transformations from one to the other, possibly stopping at some given height/iteration \cite{LICATA2011263}. Though vertical in some sense, this notion of directed-ness is considered only for terms, while in our case the main focus is for types (though one could regard types as terms of a given universe, in DTT transformations are introduced for \emph{all} pairs of terms). Being based on ($\omega$-)categories, they share some properties of morphisms, such as composition and identity, but our vertical maps are a lot simpler, and in no way higher-dimensional -- though the monad in \autoref{monad_gcwf} suggests possible extensions in this direction.

\medskip

\noindent\textit{Observational type theory.}
Observational type theory was introduced to combine the nice computational features of intensional type theory, such as termination of reductions, and propositional equality of extensional type theory, so that two functions are equal if they are equal point-wise, or ``if all observations about them agree''  \cite[\S1]{Altenkirch2006TowardsOT}. The fact that conversion rules allow to pass implicitly between definitionally equal types is extended to a mechanism that allows to pass implicitly between \emph{provably} equal types. This process is explicit and its agents are called \emph{coercions}. Coercions for a OTT work in a way that is much similar to our vertical maps, but their being originated by proofs of equality of types ensures that for a given coercion, one can always one associated to it and going in the opposite direction.

\medskip

\noindent\textit{Practical subtyping.}
Another relevant syntactic approach is that of `practical subtyping' as introduced in \cite{LepigreR19}. There, the new subtyping judgements are \emph{ternary} relations as $t\in A\subset B$, which are encoded as sorts of implications: ``if $t$ is a term of type $A$, then it is a term of type $B$''. In particular, this notion of subtyping most notably (and `practically') describes subtyping from the perspective of terms, and not types, as in our case. In particular, subtyping between types, meaning intrinsic judgements such as $A\subseteq B$, are encoded by means of choices operators, such as Hilbert's $\epsilon$ \cite[p. 26]{LepigreR19}.

\section{Examples and applications}\label{sec:3}
\subsection{Gcwfs from Lindenbaum-Tarski}\label{lt-fibration}
Consider the Lindenbaum-Tarski algebra of a given first-order theory $\mathcal{T}$ in a language $\mathcal{L}$. We take $\catof{ctx}$ to be the category where
\begin{itemize}
\item objects are lists of distinct variables $x=(x_1,\dots,x_n)$,
\item arrows are lists of substitution for variables, meaning $[t_1/y_1,\dots,t_m/y_m]=[t/y]\colon x\to y$, with $t_j$'s being $\mathcal{L}$-terms that are built on variables $x_1,\dots,x_n$,
\end{itemize}
and composition is defined by simultaneous substitution. The product of two lists $x$ and $y$ is a list $w$ with length the sum of the lengths of $x$ and $y$, and projections on $x$ and $y$ are substitutions with, respectively, the first $n$ and the last $m$ variables in $w$. Categorically and in the sense of \cite{lawvere1963functorial}, this is the free Lawvere theory on the language $\mathcal{L}$.

One can define the functor $LT_{\mathcal{T}}\colon\catof{ctx}\opp\to\catof{InfSL}$ so that to each list $x$, the category $LT_{\mathcal{T}}(x)$ has for objects equivalence classes of well-formed formulae in $\mathcal{L}$ with free variables at most those that are in $x$, and with respect to the equivalence relation induced by reciprocal deducibility in $\mathcal{T}$, $\phi \dashv \vdash_{\mathcal{T}} \phi'$. Notice that this makes our treatment \emph{proof-irrelevant}. Maps in $LT_{\mathcal{T}}(x)$ are provable consequences in $\mathcal{T}$. Composition is given by the cut rule of the calculus, and identities are tautologies. Since substitution preserves provability, $LT_{\mathcal{T}}$ can be suitably extended to a functor, and its correspondent under the Grothendieck construction \autoref{groth_equivalence} is a faithful fibration $p\colon \int LT_{\mathcal{T}}\to\catof{ctx}$.

Such a construction was first introduced with the name of \emph{(hyper)doctrine} in \cite{equalityinhd} and is thoroughly explained in \cite{kock1977doctrines} and \cite{mr_quotientcompl13}. We here show that it underlies the structure of a gcwf where terms are entailments.
\begin{example}
Call $\ee$ the category of $p$-vertical maps and commutative squares, which is again fibered over $\catof{ctx}$ -- call $e$ this fibration, $\mathsf{Cod}\colon\ee\to\int LT_{\mathcal{T}}$ the codomain functor, $\mathsf{Diag}$ the functor mapping each formula to its identity. The triple $(p,e,\mathsf{Cod}\dashv\mathsf{Diag})$ is a gcwf.
\end{example}
This gcwf is in fact \emph{not} a cwf. Here, types are formulae and terms are (unique) witnesses to entailment, meaning triples $x;\phi\vdash\psi$ where $\phi$ and $\psi$ are both formulae in the fiber over $x$. The underlying notion of subtyping actually coincides with terms.
\begin{remark}
This is an example of a more general instance, which will thoroughly discussed in \autoref{sec:4}.
\end{remark}

\subsection{Gcwfs from topos theory}
Let $\ee$ an elementary topos and $\top\colon\one\to\Omega$ its subobject classifier, then consider
\[\begin{tikzcd}[ampersand replacement=\&]
	\ee_{/\one} \&\& \ee_{/\Omega} \\
	\& \ee
	\arrow["\dom", from=1-3, to=2-2]
	\arrow["\sim"', from=1-1, to=2-2]
	\arrow[""{name=0, anchor=center, inner sep=0}, "\Sigma_\top"', curve={height=6pt}, from=1-1, to=1-3]
	\arrow[""{name=1, anchor=center, inner sep=0}, "\Delta_\top"', curve={height=6pt}, from=1-3, to=1-1]
\end{tikzcd}\]
with $\Sigma_\top=\top\circ\text{-}$ and $\Delta_\top(\phi\colon X\to \Omega)=(\text{canonical p.b. of }\phi\text{ along }\top)$ .
\begin{example}
The triple $(\sim,\dom,\Sigma_\top\dashv\Delta_\top)$ is a gcwf.
\end{example}
Compatibly with the Mitchell-Bénabou interpretation, types are (proof irrelevant) propositions, $\Delta_\top$ computes the comprehension $\{x\,|\,\phi(x)\}$, and so on -- we refer to \cite[Part D]{elephant1}, \cite[Chap. II]{lambek1988introduction} for an extensive treatment of the topic, and to \cite{low_mb} for a quick overview. In this case, the fibrations involved are discrete, hence this is really is a cwf, and there is no subtyping.

A topos is quite a general structure, so let us break down a couple of examples for it. The prototypical example of an elementary topos is $\catof{Set}$, the categories of sets and functions, with subobject classifier $\top\colon\one\to\mathsf{2}$, $\ast\mapsto 1$, which classifies subsets via their characteristic function.
\[\begin{tikzcd}[ampersand replacement=\&]
	{A_\phi} \& \one \\
	A \& {\mathsf{2}}
	\arrow["\top", from=1-2, to=2-2]
	\arrow["\phi"', from=2-1, to=2-2]
	\arrow["{i_\phi}"', hook', from=1-1, to=2-1]
	\arrow[from=1-1, to=1-2]
	\arrow["\lrcorner"{anchor=center, pos=0.125}, draw=none, from=1-1, to=2-2]
\end{tikzcd}\]
In this case terms are sets, types are functions $\phi\colon A\to \mathsf{2}$, which are equivalently subsets of $A$ -- and the reason why we usually call them \emph{propositions}. The context for each proposition $\phi$ is its domain set $A$. Let us now break down rules as in (\ref{rules-gcwf}),
\begin{gather}
(\Sigma_\top)\frac{A\vdash A}{A\vdash i_{\id_A}\ty}
\quad
(\Delta_\top)\frac{A\vdash \phi\ty}{A_\phi\vdash A_\phi}
\end{gather}
where we use the fact that $\Delta_\top(\phi)=\dom i_\phi=\{a\,|\,\phi(a)\}\subseteq A$ by definition. Notice how, again, $\Delta$ does not make the triangle commute, as, in principle, $A\neq A_\phi$.

Let us now look at a more complicated case, namely that of $\catof{Eff}$, the effective topos \cite{HYLAND1982165,2a53f234-d93d-3ee4-ba4d-a0e9700e3eab}, whose objects are sets $A$ with an $\nn$-valued equality predicate $=_A\in\mathcal{P}(\nn)^{A\times A}$, and whose morphisms are $\mathcal{P}(\nn)$-valued functional relations: we think of each subset of $\nn$ as the \emph{realizers} (in the sense of Kleene) for a given proposition. One can show that $\catof{Eff}$ is a topos with subobject classifier $\top\colon (\{\ast\},\nn)\to(\mathcal{P}(\nn), \leftrightarrow)$, where $A\leftrightarrow B:= (A\to B)\land(B\to A)$ is a pair of recursive functions mimicking bi-implication \cite[\S1]{HYLAND1982165}, and $\top(\ast,A)=[A\leftrightarrow \nn]$. In this case, terms are sets with an $\nn$-valued equality predicate, and types are subobjects of them, which can be shown to be sort of subsets with a compatible equality predicates and an additional `membership' unary predicate compatible with it.

\subsection{Gcwfs from pullbacks}
For $\ctg{C}$ with pullbacks we can define $\ctg{C}\due$ the category of arrows in $\ctg{C}$ and $\catof{Sec}(\ctg{C})$ of sections of arrows in $\ctg{C}$, meaning of pairs $(s,f)$ with $f,s\in\ctg{C}\due$ and $f\circ s =\id$. Therefore a type in context $A$ is a map $f\colon B\to A$ and a term of type $f$ is one of its sections, $s$. For a given type $f$ (map) context extension can be performed using what is usually called its `kernel pair' construction, meaning computing the following pullback,
\[\begin{tikzcd}[ampersand replacement=\&]
	B \\
	\& {K_f} \& B \\
	\& B \& A
	\arrow["f"', from=3-2, to=3-3]
	\arrow["f", from=2-3, to=3-3]
	\arrow["{f^+}", from=2-2, to=3-2]
	\arrow[from=2-2, to=2-3]
	\arrow["\lrcorner"{anchor=center, pos=0.125}, draw=none, from=2-2, to=3-3]
	\arrow["\id"', curve={height=6pt}, from=1-1, to=3-2]
	\arrow["\id", curve={height=-6pt}, from=1-1, to=2-3]
	\arrow["{!\, d(f)}"{description}, dashed, from=1-1, to=2-2]
\end{tikzcd}\]
which we can interpret as the universal $f$-induced congruence on $B$. One can always compute the section $B\to K_f$, which can be extended to a functor $K\colon\ctg{C}\due\to\catof{Sec}(\ctg{C})$.

\begin{example}
Call $U\colon\catof{Sec}(\ctg{C})\to\ctg{C}\due$ the functor mapping each pair $(s,f)$, with $s$ section of $f$, to $f$. The triple $(\cod,\cod U,U\dashv K)$ is a gcwf.
\end{example}
Actually, by \autoref{biequiv_thm}, the same can be said for any $\ctg{D}$ subcategory of $\ctg{C}\due$ closed for pullbacks, for example we can consider monomorphisms, and their sections. Notice that this is \emph{not} a cwf, as vertical maps are commutative triangles.

Let us give a couple of examples of this case, as well. We look at $\Set$ again, for a function $f\colon B\to A$, we find that $K_f$ is the set of pairs of elements of $B$ with the same image through $f$, and $B\to K_f$ is the diagonal. We again try to break down rules as in (\ref{rules-gcwf}), and get the following.
\begin{gather}
(U)\frac{A\vdash f\colon B \leftrightarrows A \colon s}{A\vdash f\colon B\to A\ty}
\quad
(K)\frac{A\vdash f\colon B\to A\ty}{B\vdash f^+\colon K_f \leftrightarrows B \colon d(f)}
\end{gather}
Again, it is important that the right adjoint $K$, the one performing context extension, does not make the triangle commute, as $B$ is not (necessarily) equal to $A$. Let us now look at subtyping judgements: morphisms in $\ctg{C}\due$ are commutative squares, vertical ones are squares such that the codomain is the identity, hence commutative triangles. The whole set of structural judgements available in this model (cf. \autoref{new_judgements}) is then the following.
\begin{center}
\adjustbox{width=\textwidth}{
\begin{tabular}{cccc}
$A \vdash f\ty$ & $A \vdash s:f$ & $A \vdash f'\leq_h f$ & $A \vdash s:_k f$ \\
&&&\\
\begin{tikzcd}[ampersand replacement=\&]
	{\dom f} \\
	A
	\arrow["f", from=1-1, to=2-1]
\end{tikzcd} & \begin{tikzcd}[ampersand replacement=\&]
	{\dom f} \\
	A
	\arrow["f", curve={height=-6pt}, from=1-1, to=2-1]
	\arrow["s", curve={height=-6pt}, from=2-1, to=1-1]
\end{tikzcd} & \begin{tikzcd}[ampersand replacement=\&]
	{\dom f'} \&\& {\dom f} \\
	\& A
	\arrow["{f'}"', from=1-1, to=2-2]
	\arrow["f", from=1-3, to=2-2]
	\arrow["h", from=1-1, to=1-3]
\end{tikzcd} & \begin{tikzcd}[ampersand replacement=\&]
	{\dom f'} \&\& {\dom f} \\
	\& A
	\arrow["{f'}", curve={height=-6pt}, from=1-1, to=2-2]
	\arrow["f", from=1-3, to=2-2]
	\arrow["k", from=1-1, to=1-3]
	\arrow["s", curve={height=-6pt}, from=2-2, to=1-1]
\end{tikzcd}\\
&&&
\end{tabular}}
\end{center}
Of course one could consider a great multitude of other categories, such as $\catof{Pos},\catof{Top}$, $\catof{Vect}_K, \catof{Grp}$, all topoi, and many others.

\subsection{On type constructors}\label{constructors}

We consider here only function types of the form $\funty(A,B)$,
where $A$ and $B$ are in the same context
(in particular, $B$ does not depend on $A$).
In this case, given a subtype $A'$ of $A$ and a subtype $B$ of $B'$
(note the different order),
one would like to conclude that $\funty(A,B)$ is a subtype of $\funty(A',B')$.
In our proposed framework, this can be expressed by requiring an action of $\funty$ on vertical arrows, which is contravariant in the first component.

In the case of a discrete gcwf
(that is, in the fibrational formulation of a natural model/cwf)
the existence of function types can be expressed by requiring the existence of functors $\funty$ and $\absty$ making the right-hand square below a pullback~\cite{awodey_2018}
\[\begin{tikzcd}
\uu {}_{\du\Delta}\!\!\times_{\du} \duu	\ar[d,"{\id \times \Sigma}"']
&	\mathrm{W}^*(\uu {}_{\du\Delta}\!\!\times_{\du} \duu)
	\ar[d] \ar[r,"\absty"] \ar[l]
&	\duu	\ar[d,"{\Sigma}"]
\\
\uu {}_{\du\Delta}\!\!\times_u \uu
&	\uu {}_u\!\!\times_u \uu
	\ar[l,"\mathrm{W}"'] \ar[r,"\funty"]
&	\uu
\end{tikzcd}\]
where the left-hand square is a pullback.
The lower functor $\funty$ simply takes two types $A$ and $B$ in the same context $\Gamma$ to the type $\funty(A,B)$.
The functor $\mathrm{W}$ is weakening of the second type with the first one:
it takes a pair $(A,B)$ of two types $A$ and $B$ in the same context $\Gamma$ to the pair $(A,(u\epsilon_A)^*B)$
where $(u\epsilon_A)^*B$ is the type $B$ weakened to the context $\Gamma.A$.
As the left-hand square is a pullback,
the domain of the functor $\absty$ consists of pairs $(A,b)$ of a type $A$ in context $\Gamma$ and a term $b$ of the weakened type $(u\epsilon_A)^*B$ in context $\Gamma.A$.
The functor $\absty$ simply maps such a pair to the term $\absty(A,b)$.
Commutativity of the square ensures that the type of $\absty(A,b)$ is $\funty(A,B)$.
As it is not relevant for our discussion,
we refer to~\cite{awodey_2018} for details on how the pullback property of the right-hand square validates both the elimination rule and the $\eta$-rule.

To extend this setting to include subtyping in the form of vertical arrows,
we need to take into account that the action of $\funty$ (and $\absty$) should be contravariant \emph{only on the vertical arrows} of the first component.
Contravariant actions of functors on some category $\ctg{B}$ are rendered by considering functors on the opposite category $\ctg{B}\opp$,
but here we want to take the opposite only of vertical arrows,
otherwise substitution, \ie the action of $\funty$ on cartesian arrows, would go in the wrong direction.
So we cannot just replace $u$ with $u\opp \colon \uu\opp \to \ctx\opp$.
Instead, we want to construct a fibration
$u\vopp \colon \uu\vopp \to \ctx$
that has the same cartesian arrows of $u$, and vertical arrows going in the opposite direction
This can be done passing through the associated pseudo-functor:
write $\mathrm{P} \colon \Fib(\ctx) \to \Psd[\ctx\opp,\Cat]$
and $\mathrm{F} \colon \Psd[\ctx\opp,\Cat] \to \Fib(\ctx)$
for the two 2-functors realising the 2-equivalence in \cref{groth_equivalence}.
Then the fibration $u\vopp$ is the image under $\mathrm{F}$
of the pseudo-functor $\mathrm{P}(u)^{\text{o}} \colon \ctx^{\text{op}} \to \Cat$ defined by $\mathrm{P}(u)^{\text{o}}(X):=(\mathrm{P}(u)(X))^{\text{op}}$.
Equivalently, $u\vopp:= \mathrm{F}((-)\opp \circ \mathrm{P}(u))$.
It follows from the last equation that $(-)\vopp$ is 2-functorial.
%

\begin{proposition}\label{funct-subtypes}
Let $(u,\du,\Sigma,\Delta)$ be a gcwf equipped with two functors $\funty$ and $\absty$ making the square below a pullback.
\[\begin{tikzcd}
\uu\vopp {}_{\du\vopp\Delta\vopp}\!\!\times_{\du} \duu
	\ar[d,"{\id \times \Sigma}"']
&	\mathrm{W}^*(\uu\vopp {}_{\du\vopp\Delta\vopp}\!\!\times_{\du} \duu)
	\ar[d] \ar[r,"\absty"] \ar[l]
&	\duu	\ar[d,"{\Sigma}"]
\\
\uu\vopp {}_{\du\vopp\Delta\vopp}\!\!\times_u \uu
&	\uu\vopp {}_{u\vopp}\!\!\times_u \uu
	\ar[l,"\mathrm{W}"'] \ar[r,"\funty"]
&	\uu
\end{tikzcd}\]
Then, in addition to the usual formation, introduction, elimination and computation rules for function types, the following rules are satisfied.
\[
\AxiomC{$\Gamma\vdash A' \leq_f A$}
\AxiomC{$\Gamma\vdash B\leq_g B'$}
\BinaryInfC{$\Gamma\vdash \funty(A,B) \leq_{\funty(f,g)} \funty(A',B')$}
\DisplayProof
\hspace{3em}
\AxiomC{$\Gamma.A \vdash b :_f B$}
\UnaryInfC{$\Gamma \vdash \absty(A,b) :_{\funty(\id_A,f)} \funty(A,B)$}
\DisplayProof
\]
\end{proposition}

\medskip

\section{The `subtyping' monad}\label{sec:4} 
We now deepen our intuition for both the properties and the features of this new construction. In particular, the process of \emph{taking into consideration} vertical maps turns out to amount to the result of the action of a particular monad. Grothendieck fibrations are known to be strongly linked to certain (co)monads, to the point of them being classified as pseudo-algebras of a given one (cf. \cite{emmenegger2023comonad} for a recent review and extension). We here present a simple monad that has the nice feature of collecting vertical maps, and show that it actually produces gcwfs out of gcwfs.

We start from the endofunctor $(\Id_{-}/\Id_{-})\colon\Fib(\bb)\to\Fib(\bb)$ computing for each fibration $p\colon\ee\to\bb$ the \emph{comma object} $(\Id_p/\Id_p)$, meaning the following (very boring) 2-limit below.
\[\begin{tikzcd}[ampersand replacement=\&]
	{(\Id_p/\Id_p)} \& p \\
	p \& p
	\arrow["\Id_p"', from=2-1, to=2-2]
	\arrow["\Id_p", from=1-2, to=2-2]
	\arrow[from=1-1, to=2-1]
	\arrow[from=1-1, to=1-2]
	\arrow[shorten <=3pt, shorten >=3pt, Rightarrow, from=1-2, to=2-1]
\end{tikzcd}\]
To make the reding a bit smoother, we denote $(\Id_p/\Id_p)$ by $(p/p)$. Unpacking this categorical construction, one finds that the new fibration $(p/p)\colon\vcomma{\ee}{\ee}\to\bb$ has in the domain triples $(f,A',A)$ such that $f\colon A'\to A$ is a $p$-vertical map, while morphisms in $\vcomma{\ee}{\ee}$ are simply commutative squares.

\begin{lemma}\label{monad_fib}
There is a monad on $\Fib(\bb)$ with endofunctor $(\Id_{-}/\Id_{-})$.
\end{lemma}
Its associated unit has components $\eta\colon p \to Tp$ sending each object to its identity, while the multiplication $\mu_p\colon TTp\to Tp$ is fiber-wise composition -- as objects in the total category of $TTp$ are squares of vertical maps, and vertical maps compose.

This simple construction we can actually extend to a gcwf, meaning we can use it to build out of a gcwf with type fibration $u$, a new gcwf with type fibration $Tu$.
\begin{proposition}\label{gcwf_to_gcwf}
Let $(u,\du,\Sigma \dashv\Delta)$ a gcwf. Then there are adjoint functors $\overline{\Sigma}\dashv\overline{\Delta}$ such that $(u/u,\Sigma/u,\overline{\Sigma}\dashv\overline{\Delta})$ is a gcwf.
\end{proposition}
If we look at the basic judgments this second gcwf is describing, we will find precisely the two introduced in \autoref{new_judgements}, so that we can reformulate our old perspective:
\begin{center}
\begin{tabular}{cc}
$\Gamma \vdash A\ty$ & $\Gamma \vdash a: A$   \\
$u(A)=\Gamma$ & $\du(a)=\Gamma,\Sigma (a) =A$ \\
&\\
$\Gamma \vdash A'\leq_f A$ & $\Gamma \vdash a:_g A$\\
 $f\colon A'\to A$, $u(f)=\id_\Gamma$ & $g\colon \Sigma a\to A$, $u(g)=\id_\Gamma$
\end{tabular}
\end{center}
\vspace{.5em}
into the newer following notation.
\vspace{.5em}
\begin{center}
\begin{tabular}{cc}
$\Gamma \vdash A\ty$ & $\Gamma \vdash a: A$ \\
$u(A)=\Gamma$ & $\du(a)=\Gamma,\Sigma (a) =A$\\
&\\
 $\Gamma \vdash A'\leq_f A$ & $\Gamma \vdash a:_g A$\\
 $(u/u)(f,A',A)=\Gamma$ & $(\Sigma/u)(g,a,A)=\Gamma$, $p_2\circ\overline{\Sigma}(g,a,A)=A$
\end{tabular}
\end{center}
\vspace{.5em}
Notice now the symmetry of the first and the second two judgments: the two new ones can be regarded as `classifiers' for new types and terms themselves, in a way such that it is always possible to recover the original theory -- one simply has to look at those vertical maps that, in particular, are identities.
\begin{definition}[The category of gcwfs]
A \emph{gcwf morphism} $(u,\du,\Sigma\dashv\Delta)\to(v,\dv,\Sigma'\dashv\Delta')$ is the data of a pair $(H,\dot{H})$, with $H\colon u\to v$, $\dot{H}\colon \du\to\dv$ fibration morphisms such that $H\Sigma =\Sigma'\dot{H}$. We denote $\gcwf$ the category of gcwfs and gcwf morphisms, and $\gcwf(\bb)$ its subcategory with fixed context category $\bb$.
\end{definition}
Again, it should be noted that gcwfs can be equipped with more interesting (and higher dimensional) structure \cite[$\S$3.2]{coraglia_phd}, but for the purposes of the present paper we are not interested in that.
\begin{theorem}\label{monad_gcwf}
The monad in \autoref{monad_fib} can be lifted to a monad over $\gcwf(\bb)$.
\end{theorem}
If a monad is an object for abstract computation, this one in particular computes `subtyping judgements'. An interesting question is now to check what objects does iterating this computation produce: the first iteration collects vertical maps, and morphisms are squares; in particular, vertical morphisms are those squares having vertical (with respect to the original type fibration) components, therefore the second iteration produced the fibration of said `all vertical' squares; here morphisms are cubes, and the vertical ones are those where all `connecting' maps are vertical (again, with respect to the original type fibration), and so on.

\section{Future work} 
Of course the applications to constructors can be taken much further than what is done in \autoref{constructors}, and considering more complex ones can lead to a new extension of the present theory. Our intention was to introduce the `vertical maps' perspective, so that whoever gets interested in categorical models for type theory is motivated to consider the benefit of including non-discrete fibrations -- usually, all in all, at little additional cost.

We conclude this exposition with a take that was suggested to the first author by F. Dagnino, and which we believe might be of some interest: as a fibration can be regarded as an internal category in the category of discrete fibrations, it seems like the process of going from a theory \emph{without} a notion of subtyping to a theory \emph{with} subtyping is close to that of considering a category {internal to} another one. In this sense, our work is perhaps not best interpreted as a generalization of some known models, but as a structure internal to them.

\bibliography{thebib}

\appendix
\section{Proofs}

\begin{proof}[Proof of \autoref{strrules}]
The first two rules, Subsumption \textit{(Sbsm)} and Transitivity \textit{(Trans)}, simply follow from composition of vertical arrows. Substitution \textit{(Sbst)} and Weakening \textit{(Wkn)} are a bit trickier and make use of, precisely, the substitutional part of the structure, namely it being a fibration. We begin with Weakening, as it is a bit easier: consider that the premise of the rule is given by maps and objects as below.
\[\begin{tikzcd}[ampersand replacement=\&]
	\& {A'} \\
	\& A \& \uu \\
	{\Gamma.B} \& \Gamma \& \bb
	\arrow["{u\epsilon_B}", from=3-1, to=3-2]
	\arrow["u", from=2-3, to=3-3]
	\arrow["f", from=1-2, to=2-2]
\end{tikzcd}\]
Since $B$ is a type in context $\Gamma$, we can compute the counit $\epsilon\colon\Sigma\Delta B\to B$, which we know by (\ref{rules-gcwf}) that acts as weakening, in particular producing the context extension $u\epsilon_B\colon\Gamma.B\to\Gamma$. In the fiber over $\Gamma$, not only do we have $B$, but we also have a vertical map $f\colon A'\to A$. Applying the reindexing $(u\epsilon_B)^*$ to the three of them yields a new vertical map over $\Gamma.B$, hence the judgement $\Gamma.B\vdash (u\epsilon_B)^*A'\leq_{(u\epsilon_B)^*f} (u\epsilon_B)^*A$. In \autoref{strrules} we have actually written $A$ for $(u\epsilon_B)^*A$, at it is customary for weakening rules -- and as, in fact, it appears in (\ref{rules-gcwf}). All in all, the Weakening rule only tells us that subtyping is preserved through weakening.

Now let us get to Substitution. In order to prove it, we first need to recall how $B[a]$ is computed in a traditional natural model (cf. \cite[\S2.1]{awodey_2018}): for given $\Gamma\vdash a:A$ and $\Gamma.A\vdash B$, $B[a]$ is the result of the reindexing $B$ along the (image of the) unit $\eta_a$. (This is not precisely how the original paper presents it, though it is equivalent to that: see \cite[\S2.3.4]{coraglia_phd} for a detailed discussion on the matter.) In our setting, the premises of \textit{(Sbst)} start from the following.
\[\begin{tikzcd}[ampersand replacement=\&]
	{\Sigma a} \& {\Sigma\Delta\Sigma a} \&\& {\Sigma a} \\
	A \&\& {\Sigma\Delta A} \& A \& \uu \\
	\Gamma \& {\Gamma.\Sigma a} \& {\Gamma.A} \& \Gamma \& \bb
	\arrow["{\du\Delta g}", from=3-2, to=3-3]
	\arrow["u", from=2-5, to=3-5]
	\arrow["{\du\eta_a}", from=3-1, to=3-2]
	\arrow["{u\epsilon_{\Sigma a}}", from=3-3, to=3-4]
	\arrow["g", from=1-4, to=2-4]
	\arrow[from=1-2, to=1-4]
	\arrow[from=2-3, to=2-4]
	\arrow[from=1-2, to=2-3]
	\arrow[dashed, from=2-1, to=2-3]
	\arrow[from=1-1, to=2-1]
	\arrow[from=1-1, to=1-2]
\end{tikzcd}\]
Notice that, since $g$ is vertical, and the top horizontal compositions in the top diagram is an identity due to triangle identities of $\Sigma\dashv\Delta$, by cartesianness of the counit $\epsilon_A\colon\Sigma\Delta A\to A$ there is a unique map making the top diagram commute. Now, each `external' side in the top diagram is mapped to the identity through $u$, meaning we can repeat Awodey's argument, and the required vertical arrow is the result of reindexing along the (image of the) unit -- post-composed with $\du\Delta g$. Notice that $a$ is only `allowed' to interact with $B,B'$ through $g$, which in turn does nothing to contexts, so that we are in our right to write, as customary, $B[a]$ for $(\du(\Delta g\circ\eta_a))^*B$.
\end{proof}
\begin{remark}
It might not seem like it, but this last proof makes heavy use of the universal property of the comma object $(u/u)$, and it could be in fact entirely proved using that profusely. We point the interested reader to \cite[\S3.5]{coraglia_phd} for the corresponding alternative proof.
\end{remark}

\begin{proof}[Proof of \autoref{leftfaith}]
Each component of the unit in a gcwf is a monic arrow.
Indeed, let $f,g \colon a \to b$ in $\duu$ be such that $\eta_b f = \eta_b g$.
It follows that
\[
\du f = (u\epsilon_{\Sigma b})(\du \eta_b)(\du f)
= (u\epsilon_{\Sigma b})(\du \eta_b)(\du g) = \du g
\]
and, in turn, that $f = g$ since $\eta_b$ is cartesian. The left adjoint $\Sigma$ is then faithful.

In fact, it is easy to see that $\Sigma$ induces a bijection
\begin{equation}\label{jdtt-leftfaith-iso}
\begin{tikzcd}[column sep=4em]
\duu(a,b)	\ar[r,"\sim"]
&	\{ f \in \uu(\Sigma a, \Sigma b) \mid (\Sigma\eta_b)f = (\Sigma\Delta f)(\Sigma\eta_a) \}.
\end{tikzcd}
\end{equation}
Since $\eta_b$ is cartesian,
the counter-image of $f$ is the only arrow $g$ in $\duu(a,b)$ over $uf$
such that $\eta_b g = (\Delta f) \eta_a$.
\end{proof}

\begin{proof}[Proof of \autoref{funct-subtypes}]
This is straightforward, by reading the action of $\funty$ on vertical arrows
and unfolding the definitions of the judgements involved as in \autoref{new_judgements}.
\end{proof}

\begin{proof}[Proof of \autoref{gcwf_to_gcwf}]
We begin by describing what the fibration of terms $(\Sigma/u)$ does: it simply collects pairs $(g,a,A)$ such that $g\colon\Sigma a\to A$ is a $u$-vertical map, and sends them to their underlying context. Its vertical maps are pairs of vertical maps (respectively $\du$- and $u$-vertical) fitting in appropriate squares, and its cartesian maps are pairs of cartesian maps (again, respectively, with reference to $\du$ and $u$).

The required typing functor, then, is simply $\overline{\Sigma}$ sending each triple $(g,a,A)$ to $(g,\Sigma a,A)$, which is cartesian because $\Sigma$ is. Describing its right adjoint requires a little more effort: for a triple $(f,A',A)$ in $(u/u)$ one considers its image through $\Delta$, then its vertical-cartesian factorization system (cf. \autoref{vertcart_facsys}), then the image of the vertical portion of it through $\Sigma$ to get back to $\uu$.
\[\begin{tikzcd}[ampersand replacement=\&]
	{A'} \&\& {\Delta A'} \&\&\& {\Sigma\Delta A'} \&\& {A'} \\
	A \&\& {a_f} \& {\Delta A} \&\& {\Sigma a_f} \& {\Sigma\Delta A} \& A
	\arrow["f", from=1-1, to=2-1]
	\arrow["{(\Delta f)^v}"', from=1-3, to=2-3]
	\arrow["{(\Delta f)_c}"', from=2-3, to=2-4]
	\arrow["{\Delta f}", from=1-3, to=2-4]
	\arrow["{\epsilon_{A'}}", from=1-6, to=1-8]
	\arrow["{\epsilon_A}"', from=2-7, to=2-8]
	\arrow["f", from=1-8, to=2-8]
	\arrow["{\Sigma\Delta f}", from=1-6, to=2-7]
	\arrow["{\overline{\Delta}f=\Sigma(\Delta f)^v}"', from=1-6, to=2-6]
	\arrow["{\Sigma(\Delta f)_c}"', from=2-6, to=2-7]
\end{tikzcd}\]
Cartesianness of $(\Delta f)_c$ can be used to provide functoriality. Let us now show how the two functors form an adjoint pair by means of the universal property of its counit, which is actually depicted on the right square above. For any triple $(g,b,B)$ in $(\Sigma/u)$ and morphism $(h,k)\colon \overline{\Sigma}(g,b,B)\to (f,A',A)$ we must show that there is a unique pair $(m,n)$ such that $\overline{\Sigma}(m,n)\colon \overline{\Sigma}(g,b,B)\to \overline{\Sigma}\overline{\Delta}(f,A',A)$ makes the obvious triangle commute. The universal property of the counit $\epsilon_A'$ yields a unique $m\colon b\to \Delta A'$ so that $h=\epsilon_{A'}\circ m$,
\[\begin{tikzcd}[ampersand replacement=\&]
	{\Sigma b} \\
	B \&\& {\Sigma\Delta A'} \&\& {A'} \\
	\&\& {\Sigma a_f} \& {\Sigma\Delta A} \& A \\
	\Theta \\
	\&\& {\Gamma.A'} \& {\Gamma.A} \& \Gamma
	\arrow["{\epsilon_{A'}}", from=2-3, to=2-5]
	\arrow["{\epsilon_A}"', from=3-4, to=3-5]
	\arrow["f", from=2-5, to=3-5]
	\arrow["{\Sigma\Delta f}", from=2-3, to=3-4]
	\arrow["{\overline{\Delta}f=\Sigma(\Delta f)^v}"', from=2-3, to=3-3]
	\arrow["{\Sigma(\Delta f)_c}"', from=3-3, to=3-4]
	\arrow["g"', from=1-1, to=2-1]
	\arrow["h"{pos=0.1}, curve={height=-6pt}, from=1-1, to=2-5]
	\arrow["k"'{pos=0.1}, curve={height=-6pt}, from=2-1, to=3-5]
	\arrow["{\Sigma(!)}"', dashed, from=1-1, to=2-3]
	\arrow[from=5-4, to=5-5]
	\arrow[curve={height=-6pt}, from=4-1, to=5-5]
	\arrow[from=5-3, to=5-4]
	\arrow["{u\Sigma(!)}"', from=4-1, to=5-3]
\end{tikzcd}\]
hence a map filling the bottom diagram to a commuting triangle. By cartesianness of $\epsilon_A\circ\Sigma(\Delta f)_c$ there is a unique map $n\colon B\to\Sigma a_f$ making the `other' triangle commute as well.
\end{proof}

\begin{proof}[Proof of \autoref{monad_gcwf}]
Let us denote $(T,\eta,\mu)$ the desired monad. On objects it of course acts as $T(u,\du,\Sigma\vdash\Delta)=((u/u), (\Sigma/u), \overline{\Sigma}\dashv\overline{\Delta})$ as in \autoref{gcwf_to_gcwf}. The image of a morphism $(H,\dot{H})$ is actually simply the action of $H$, and it provides a commutative square because $H\Sigma=\Sigma'\dot{H}$. Unit is again induced by identity, and multiplication by composition.
\end{proof}
\end{document}

%% file: header.tex
\usepackage{amsthm,amsmath,amssymb,amsfonts}
\usepackage[utf8]{inputenc}
\usepackage{verbatim}
\usepackage{marvosym}
\usepackage{stackrel}
\usepackage{graphicx}
\usepackage{bbold}
\usepackage{stmaryrd}
\usepackage{lipsum}
\usepackage{svg}
\usepackage{bbm}
\usepackage{caption}
\usepackage{enumitem}
\usepackage{epigraph}
\usepackage{color}
\usepackage{microtype}
\usepackage{relsize}
\usepackage{amsfonts}
\usepackage{adjustbox}
\usepackage{subfig}
\usepackage{framed}
\usepackage{hyperref}
\hypersetup{
	colorlinks = true,
	linkbordercolor = {red},
	linkcolor ={teal},
	anchorcolor = {pink},
	citecolor =  {orange},
	filecolor = {teal},
	menucolor = {teal},
	runcolor =  {teal},
	urlcolor = {teal},
}
\usepackage[capitalize]{cleveref}
\usepackage{bussproofs} 
\usepackage{tcolorbox} 
\usepackage{csquotes}

\usepackage{tikz-cd}
\tikzcdset{scale cd/.style={every label/.append style={scale=#1},
		cells={nodes={scale=#1}}}}
\newcounter{nodemaker}
\tikzset{%
	symbol/.style={%
		draw=none,
		every to/.append style={%
			edge node={node [sloped, allow upside down, auto=false]{$#1$}}}
	}
}
\usepackage{quiver}

\DeclareFontFamily{U}{min}{}
\DeclareFontShape{U}{min}{m}{n}{<-> udmj30}{}
\DeclareUnicodeCharacter{3088}{\yo}
\makeatletter
\newcommand{\big@doubleop}[1]{%
	\DOTSB\mathop{\mathpalette\big@doubleop@aux{#1}}\slimits@
}
\newcommand*{\doublerightarrow}[2]{\mathrel{
		\settowidth{\@tempdima}{$\scriptstyle#1$}
		\settowidth{\@tempdimb}{$\scriptstyle#2$}
		\ifdim\@tempdimb>\@tempdima \@tempdima=\@tempdimb
		\mathop{\vcenter{
				\offinterlineskip\ialign{\hbox to\dimexpr\@tempdima+1em{##}\cr
					\rightarrowfill\cr\noalign{\kern.5ex}
					\rightarrowfill\cr}}}\limits^{\!#1}_{\!#2}}}
\newcommand*{\triplerightarrow}[1]{\mathrel{
		\settowidth{\@tempdima}{$\scriptstyle#1$}
		\mathop{\vcenter{
				\offinterlineskip\ialign{\hbox to\dimexpr\@tempdima+1em{##}\cr
					\rightarrowfill\cr\noalign{\kern.5ex}
					\rightarrowfill\cr\noalign{\kern.5ex}
					\rightarrowfill\cr}}}\limits^{\!#1}}}
\makeatother

\newcommand{\hirayyyy}{\text{\usefont{U}{min}{m}{n}\symbol{'110}}}
\DeclareFontFamily{U}{min}{}
\DeclareFontShape{U}{min}{m}{n}{<-> dmjhira}{}

\newcommand{\yo}{\hirayyyy}

\newtheorem*{thm*}{Theorem}

\newtheorem{theorem}{Theorem}[]
\newtheorem{definition}[theorem]{Definition}
\newtheorem{example}[theorem]{Example}
\newtheorem{lemma}[theorem]{Lemma}
\newtheorem{remark}[theorem]{Remark}
\newtheorem{proposition}[theorem]{Proposition}
